\documentstyle[12pt]{article}
 1

\begin{document}

\begin{flushright}
{CERN-TH.7366/94\\
CRETE-TH.07/94\\
IOA.306/94}\\
\end{flushright}

\vspace*{3mm}
\begin{center}
{\bf   LARGE TOP AND BOTTOM YUKAWA}\\
{\bf    COUPLINGS IN MINIMAL SUPERSYMMETRY}\\
\vspace*{1cm}
{\bf E. G. Floratos}\\
{\it  Physics Department} \\
{\it  University of Crete} \\
{\it Iraklion, Crete}\\
{\it Greece} \\
\vspace*{0.3cm}
{\bf and}\\
\vspace*{0.3cm}
{\bf G. K. Leontaris{\footnote{Permanent address:
 Theoretical Physics Division, Ioannina University,
 GR-45110 Greece}}}\\
{\it CERN} \\
{\it Theory Division} \\
{\it 1211 Geneva 23} \\
 {\it Switzerland}\\
\vspace*{0.5cm}
{\bf ABSTRACT} \\
\end{center}
\noindent

We present analytic expressions for the top and bottom Yukawa couplings
in the context of the minimal supersymmetric standard model
when both couplings $h_{t,0},h_{b,0}$ are large at the unification
scale. For sufficiently large $h_{t,0},h_{b,0}$, i.e. within $\sim
(20-25)\% $ close to their fixed point values, using as input the
central value of the bottom mass $m_b(m_b)=4.25GeV$, we find that the
top mass lies in the range $m_t \approx (174-178)GeV$, while $tan\beta
\approx (55-58)$. Implications on the evolution of the scalar masses
and the radiative symmetry breaking scenario are discussed.

\vspace*{1cm}
\noindent
\begin{flushleft}
{CERN-TH.7366/94\\
CRETE-TH.07/94\\
IOA.306/94}
\\
July 1994
\end{flushleft}
\thispagestyle{empty}
\vfill\eject
\setcounter{page}{1}


 \noindent

It is widely believed that the minimal supersymmetric standard model
 (MSSM)  is the most natural extention of the standard theory of strong
and electromagnetic interactions.  Furthermore, the MSSM can be naturally
embedded  in all unified supergravity and superstring  constructions.
Detailed calculations $\cite{calc}$ taking into account the most recent
data on the low energy  values of gauge couplings and other measurable
parameters, have shown that the above theoretical expectations are
correct when supersymmetry
 breaks at the order of $1 TeV$ and the unification of gauge couplings
takes place at the scale of $10^{16} GeV$ provided that the MSSM fermion
and higgs content is used. It was therefore recently realized
$\cite{real}$ that it is time to take a  step further and, - in addition
to the gauge coupling unification - explore  the nature of Yukawa
couplings which in the MSSM are treated as free parameters.   Grand
Unified Theories $\cite{Theo}$ predict various relations among them,
depending on the precise group chosen by the particular unification
scenario.  Furthermore, String theories relate the fate of Yukawa
couplings with that of the gauge constants.   It is expected that some of
them might be of the order of the common gauge coupling constant at the
unification point.  In fact, in all string scenaria  there appears a
hierarchical  form of the quark and lepton Yukawa couplings, due to
additional $U(1)$ symmetries which usually allow only one generation of
fermions to receive masses from the trilinear superpotential terms.
Concentrating on the quark sector, this would mean that the only tree
level Yukawa couplings which may be large and comparable to the unified
gauge coupling, are $h_t$ and $h_b$.  Knowing their initial values at the
GUT scale, we may determine their low energy values by evolving them down
using the renormalization group equations.  This has been done in many
recent works$\cite{work,grou,real}$ using numerical methods.  In the
present letter, we wish to present analytic forms for the above couplings,
when both are large at the unification scale.   Particular solutions as
in the interesting case of $h_{t,0}\approx h_{b,0}$ have already appeared
in the literature\cite{grou,dl}. Such forms might prove extremely useful,
for example, in the calculation of the scalar masses and the Higgs mass
parameters. In particular, the latter play  a very important role in the
radiative symmetry breaking scenario $\cite{scen}$. In particular in the
large $tan\beta $ scenario where  both couplings are large, the higgs
mass-squared parameters may both be driven negative and the stability
of the neutral higgs potential will be questionable. Therefore, analytic
forms for the mass  parameters can simplify the analysis of the
minimization conditions and provide a better insight of the role of the
Yukawa coupling contributions to the scalar masses.

In the following, we will assume that only  $h_t,h_b$, Yukawa couplings
are large. We will assume that  all other Yukawa couplings, including
$h_{\tau}$  are small and will ignore them. Thus, we do not implement
the minimal GUT relation $h_{\tau ,0}\equiv h_{b,0}$ at the GUT scale.
We should point out that this happens  very often in the case of String
GUTs (as in the case of the flipped $SU(5)$ $\cite{SU(5}$), and therefore
it is of particular  interest. Nevertheless, our solutions are still a
good approximation even in the case  of appreciable initial $h_{\tau ,0}$
values.

Ignoring all other Yukawa couplings, the coupled differential system
of $h_t-h_b$ couplings is written as follows,
\begin{eqnarray}
\frac{d}{dt} h^2_t &=&
\frac{1}{8\pi^2}
 \Big\{6h^2_t + h^2_b - G_Q \Big\} h^2_t \label{eq:1} \\
\frac{d}{dt} h^2_b &=& \frac{1}{8\pi^2}
\Big\{h^2_t + 6h^2_b - G_B \Big\} h^2_b \label{eq:2}
\end{eqnarray}
with
\begin{equation}
G_Q = \sum^{3}_{i=1} c^i_Q g^2_i,\qquad  G_B = \sum^{3}_{i=1} c^i_B g^2_i
\label{eq:3}
\end{equation}
where $c^i_Q = \Big\{\frac{16}{3}, 3, \frac{13}{15}\Big\}$
$\qquad c^i_B = \Big\{\frac{16}{3}, 3, \frac{7}{15}\Big\}$ for $SU(3),
SU(2)$ and $U(1)$ respectively. In order to solve
(\ref{eq:1}-\ref{eq:2}), we choose first to absorbe  the gauge
coefficients $G_Q, G_B$ by making the transformations:
\begin{equation}
 h^2_t = \gamma^2_Q x \qquad   h^2_b =
\gamma^2_B y
\label{eq:4}
\end{equation}
where
\begin{equation}\gamma^2_I = \exp \Big\{\frac{-1}
{8\pi^2} \int^t_{t_0}  G_I({t^{\prime}})
dt^{\prime}\Big\}, \qquad I = Q,B    \label{eq:5}
\end{equation}
with $t_0 = ln M_{GUT}$.  Then, by observing
that $\gamma_Q \approx \gamma_B$, since their only
difference is only a small coefficient in the $U(1)$ factor, it
can be seen easily that the D.E's (\ref{eq:1},\ref{eq:2}),
in the approximation $\gamma_Q = \gamma_B$
transform as  follows
\begin{eqnarray}
 \frac{d}{dt} x &=& \frac{1}
{8\pi^2} \gamma^2_Q  \Big\{6x+y\Big\}x
\label{eq:6}
\end{eqnarray}
\begin{eqnarray}
\frac{d}{dt} y &=& \frac{1}
{8\pi^2} \gamma^2_Q  \Big\{x+6y\Big\}y \label{eq:7}
\end{eqnarray}
This last coupled system can give a solution for $x$ in terms
of $y$ and vice versa, leading to the algebraic equation

\begin{equation}
\Big(\frac {x-y} {x_0-y_0}\Big)^7 = \big(\frac{xy}
{x_0y_0}\big)^6  \label{eq:8}
\end{equation}
Furthermore substructing (\ref{eq:6}) from (\ref{eq:7}), we
may obtain the following D.E.
\begin{equation}
\frac{d}{dt} (x-y) = \frac{6}{8\pi^2} \gamma^2_Q  (x+y) (x-y)
\label{eq:9}
\end{equation}
while $(x+y)$ can be substituted from (\ref{eq:8}) to give a
differential equation for the difference $\omega = x-y$ of the
form
\begin{equation}
\frac{d} {dt} (x-y) = \frac{6}
{8\pi^2} \gamma^2_Q  (x-y) \sqrt{k_0(x-y)^{7/6} + (x-y)^2}
\label{eq:10}
\end{equation}
where the parameter $k_0 = 4x_0y_0/(x_0-y_0)^{7/6}$ depends on the
initial conditions $x_0\equiv h_{t,0}^2$ and $y_0\equiv h_{b,0}^2$.

In order to solve eqn.(\ref{eq:9}) we make the following transformation

\begin{equation}
u=\frac{k_0}{(x-y)^{5/6}}\equiv \frac{k_0}{{\omega}^{5/6}}\, ,
\qquad d\,{\tilde I}=\frac{6}{8\pi^2}\gamma_Q^2 d\,t
\end{equation}
 Eqn.(\ref{eq:9}) can be put in the form

\begin{equation}
\frac{u^{1/5}d\,u}{\sqrt{1+u}}=-\frac{5}{6}k_0^{6/5}d\,{\tilde I}
\end{equation}
which can be integrated to give the solution in terms of Hypergeometric
functions ${}_2F_1(a ,b ,c; z)$, i.e.
\begin{equation}
u^{7/10}{}_2F_1(\frac{1}{2},\frac{-7}{10},\frac{3}{10},\frac{1}{-u},)
-u_0^{7/10}{}_2F_1(\frac{1}{2},\frac{-7}{10},\frac{3}{10},
\frac{1}{-u_0},) =\frac{7}{12}k_0^{12/10}\tilde I(t)
\end{equation}
As $a+b-c=-1/2 <0$ and $z=-1/u$, the above solution is valid
in the entire circle $|z|=1$, i.e. as long as $|u|\ge 1$.
Once the initial values, $x_0,y_0$ are chosen,
the last equation can determine the value of the difference $\omega=x-y$.
($\tilde I(t)$ can be calculated for any given scale $\mu$.)
Then, $x$ and $y$ (and therefore $h_t$ and $h_b$), can be found with the
use of Eq(\ref{eq:8}). One finds
\begin{eqnarray}
x=\frac{1}{2}\omega (1+\sqrt{1+k_0\omega^{-5/6}})\\
y=\frac{1}{2}\omega (-1+\sqrt{1+k_0\omega^{-5/6}})
\end{eqnarray}
In the following, we
choose to distinguish two separate cases, depending on the value
of $k_0$ where the results exhibit a rather interesting simplicity.

{\it i)}
$k_0<<1$.  In this case (\ref{eq:9}) is easily integrated
to give
\begin{eqnarray}
x = y +  \frac{x_0-y_0}{1-\frac{6}{8\pi^2} (x_0-y_0)I(t)}
\equiv y + Q(t)
\label{eq:11}
\end{eqnarray}
with $I(t) \equiv \int^t_{t_0} \gamma^2_Q (t^\prime)dt^\prime$.
Substitution of (\ref{eq:11}) into (\ref{eq:3}) and
integration leads to the result
\begin{equation}
h^2_t = \gamma^2_Q h^2_{t,0} \frac{q(t) }
{1- 7h^2_{t,0} J(t)} \label{eq:12}
\end{equation}
with
\begin{equation}
q(t) = exp\Big[-\frac{1}{8\pi^2 }\int^t_{t_0} \gamma^2_Q
(t^{\prime}) Q(t^{\prime })dt^{\prime }\Big] \label{eq:13}
\end{equation}
and
\begin{equation}
J(t) = \frac{1}{8\pi^2 }\int^t_{t_0} \gamma^2_Q (t^{\prime})
q(t^{\prime })dt^{\prime } \label{eq:14}
\end{equation}
Equation (\ref{eq:7}), can also be integrated in the same way
resulting to a similar formula for the bottom quark mass.

{\it ii)} $k_0\gg 1$. In this case we can ignore the second
term in eq.(\ref{eq:9}).  A straightforward integration then
gives
\begin{eqnarray}
x = y +  \frac{x_0-y_0}{(1-\frac{7}{8\pi^2} \sqrt{x_0y_0}I(t))
^{\frac{12}{7}}}
\equiv y + \Omega (t)
\label{eq:15}
\end{eqnarray}
Returning back to the initial differential system, we may obtain
the formulae for the top and bottom Yukawa couplings.
For the top-quark we obtain
\begin{equation}
h^2_t = \gamma^2_Q  h^2_{t,0}\frac{p(t) }
{1- 7h^2_{t,0} L(t)} \label{eq:16}
\end{equation}
while for the bottom
\begin{equation}
h^2_b = \gamma^2_B h^2_{b,0} \frac{p(t)^{-1} }
{1- 7h^2_{b,0} N(t)} \label{eq:17}
\end{equation}
where
\begin{eqnarray}
p(t) = exp\Big[-\frac{1}{8\pi^2 }\int^t_{t_0} \gamma^2_Q
(t^{\prime}) \Omega (t^{\prime })dt^{\prime }\Big] \label{eq:18}
\\
L(t) = \frac{1}{8\pi^2 }\int^t_{t_0} \gamma^2_Q (t^{\prime})
p(t^{\prime })dt^{\prime }
\label{eq:19}
\\
N(t) = \frac{1}{8\pi^2 }\int^t_{t_0} \gamma^2_Q (t^{\prime})
p(t^{\prime })^{-1}dt^{\prime }
\label{eq:20}
\end{eqnarray}

In order to compare our analytic results with numerical methods, we
have also solved the RGEs (\ref{eq:1}-\ref{eq:2}) numerically for some
particular cases. As an example we present here the results obtained
for initial value of the top Yukawa coupling close to its fixed point,
i.e. $h_{t,0}=3.5$ and $h_{b,0}=1.5$. The numerical solution gives
 $h_t=1.02$ and $h_b=0.955$ at the scale $m_t\approx 170GeV$, while the
analytic expressions obtained above give  $h_t=1.025$ and  $h_b=0.957$.
We have checked that this accuracy holds for all the regions of validity
of the above analytic expressions.

In Table I, we present values of the top mass, when both
 $h_{t,0}$ and $h_{b,0}$ couplings are large. In particular, we choose
two characteristic values of the top-Yukawa coupling, (one of them very
close to its fixed point) and calculate $m_t$ (running mass) and
$tan\beta $, assuming a central value for the bottom quark, i.e.
 $m_b(m_b) =  n_b \frac{\upsilon} {\sqrt{2}} h_b \cos{\beta} =4.25 GeV$.
Here $ n_b$ includes the running from the scale $\sim m_t$ down to
the scale $m_b$ and is taken to be  $ n_b\approx 1.4$.

 A very interesting inference from this table is
that the top quark has a mass around $175$ GeV which is close to the
central value found by CDF $\cite{CDF}$, while $tan\beta  \approx
(55-58)$.  We note however, that
sparticle exchange corrections on $m_b$\cite{aras,grou}
or thresholds and other uncertainties\cite{unce}
may result to small $m_t$ corrections for specific  $h_{t,0},h_{b,0}$
regions.


\begin{center}{
\vglue 0.4cm
\begin{tabular}{||c|c|c|c||}    \hline \hline

$h_{t,0}$ &$h_{b,0}$ &$m_t$  &$tan{\beta}$\\ \hline \hline

$2.5$ &$  1.75$  &$175.6$ &$ 56.25$\\  \hline
      &$  2.00$  &$175.0  $ &$ 56.73$\\  \hline
      &$  2.25$  &$174.5$ &$ 57.07$\\  \hline
      &$  2.50$  &$174.0  $ &$ 57.32$\\  \hline
$3.5$ &$  1.75$  &$177.3$ &$ 54.85$\\  \hline
      &$  2.00$  &$176.6$ &$ 56.26$\\  \hline
      &$  2.25$  &$176.1$ &$ 56.65$\\  \hline
      &$  2.50$  &$175.7$ &$ 56.93$\\  \hline
      &$  3.00$  &$175.0  $ &$ 57.31$\\  \hline \hline
\end{tabular}
\vglue 0.2cm
{\bf Table I.} {\it $m_t$ and $\tan{\beta}$ predictions for two
initial values of $h_{t,0}$ and various $h_{b,0}$-Yukawa
couplings, using expressions (\ref{eq:16}) and (\ref{eq:17}), while
fixing $m_b  = n_b \frac{\upsilon} {\sqrt{2}} h_b \cos{\beta} = 4.25 GeV$,
with  $n_b\approx 1.4$ being the renormalization group factor from
$\mu \sim m_t$ down to $\mu = m_b$}}
\end{center}

Analytic solutions for the large $h_{b,t}$ Yukawa couplings are of
particular interest in supersymmetric theories.  The masses of the
 higgses responsible for the electroweak breaking as well as the scalar
masses of the third generation receive large negative contributions when
$h_b,h_t$ are also large. In the radiative electroweak symmetry breaking
scenario $\cite{scen}$, one of the higgs mass-squared parameters should
become negative.  This is in fact possible because of the large negative
Yukawa corrections.  However, when both Yukawa couplings are large, both
higgs mass parameters receive large contributions.
Thus, when dealing with the RGE of the scalar masses
it is very useful to have analytic expressions for the Yukawa
functions which appear in the role of  scale-dependent coefficients
in the Differential Equations of the former.

In the following, we discuss briefly the effects of $h_t, h_b$ couplings
to  the Higgs and scalar quark masses of the third generation,
disentagling the coupled differential system of them and reducing it down
to a simple differential equation of second order which may be solved
either numerically or by standard analytic mathematical methods.

We start first by reminding the reader with the exact solution of the
scalar masses  $m_{{\tilde t}_L}\equiv   \tilde m_{U_1}$,
$m_{\tilde t_R} =\tilde m_{U_2}$, and $m_{{\tilde H}_2} \equiv \tilde
m_{U_3}$ in the case where  $h_t\gg h_b$, and $\tan{\beta} \simeq {\cal
O}(1)$.  It has been found in this case that the above scalar masses are
given by a simple formula  $\cite{leo}$
\begin{eqnarray}
\tilde m^2_{U_n} =  m^2_0 + C_{U_n}(t) m^2_{1/2} - n \delta^2_{\tilde m}(t)
- n \delta^2_A(t)
\label{eq:21}
\end{eqnarray}
 where, $\delta^2_A(t)$ has been estimated to be much smaller than
 $\delta^2_{\tilde m}$ and can be ignored for the present purposes, while:
\begin{eqnarray}
\delta_m^2(t)=\left(\frac{m_t(t)}{2\pi
 v\gamma_Q(t) \sin\beta}\right)^2
\times (3m_0^2I(t)+m_{1/2}^2 J_0(t))\label{eq:22}
\end{eqnarray}
where $I(t)$ has already been defined in solving Eq.(\ref{eq:11}), while
\begin{equation}
J_0(t)  =\sum^3 _{n=1}  \int^t_{t_0} C_{Un}(t^\prime)
\gamma^2_Q(t^\prime) dt^\prime \label{eq:23}
\end{equation}

The coefficients $C_{U_n}$ can be found in the literature
  $\cite{scen,work}$. Notice that in the derivation of the above, it was
assumed that $h_{t,0} \gg h_{b,0}$ while $h_{b,0}$ was ommitted from the
equations.  However, in the case where $h_{b,0}\sim h_{t,0}$ this is no
longer valid.    Since both Yukawa couplings are $h_{t/b,0} \sim {\cal
O}(1)$, both higgs mass-squared parameters $m_{H_1}^2, m_{H_2}^2$
receive large negative contributions and  play a very important role  in
the stability of the neutral higgs potential.  This can be easily seen
from the minimization conditions  $\frac{{\partial {\cal V}_H}}{{\partial
v_i }}=0$, where $ v_i \equiv <H_i>$, which result to well known equations
\begin{eqnarray}
\frac{1}{2}M_Z^2&=&\frac{{\mu_1^2-\mu_2^2\tan ^2\beta }
}{{\tan ^2\beta -1}}  \label{eq:24} \\
\frac{1}{2}sin 2\beta&=&-\frac{m_3^2}{\mu_1^2+\mu_2^2}
\label{eq:25}
\end{eqnarray}
where  $\mu_i^2=m_{H_i}^2+\mu^2
+\sigma_i^2$, with $\sigma_i^2$ being the corrections$\cite{corr}$ to
the Higgs potential from the one-loop contributions and $\mu$ the higgs
mixing term. Therefore, these contributions should  be calculated with
great care. Analytic expressions, if possible, may be extremely useful
in the minimization conditions of the potential.

In the following, we use the same techniques as in ref.${\cite{leo}}$
to calculate the Higgs and scalar masses in the case $tan\beta \gg 1$,
or equivalently when $h_{b,0},h_{t,0}$ are large. The relevant RGEs
can be found in the literature$\cite{work,scen}$. We define
\begin{eqnarray}
U=m_{U_1}^2+m_{U_2}^2+m_{U_3}^2\label{eq:26}\\
D=m_{D_1}^2+m_{D_2}^2+m_{D_3}^2\label{eq:27}
\end{eqnarray}
where $m_{U_i}$ have been defined previously while
$m_{{\tilde b}_L}\equiv   \tilde m_{D_1}$,  $m_{\tilde b_R} =\tilde
m_{D_2}$, and $m_{{\tilde H}_1} \equiv \tilde m_{D_3}$.
By recalling the same arguments used in the solution of scalar masses
for the case $h_{t,0}\gg h_{b,0}$ we can conclude that the contribution
of the trilinear parameters $A_U, A_D$ do not play important role in the
final solutions for the scalar masses in the present case too. Therefore,
to simplify the subsequent analysis, we drop  $A_U, A_D$ terms  (the
extention of the solution to the most general case is straightforward).
It is easily  then observed that one can write the equations for the sums
of scalar  masses  in the following form:
\begin{eqnarray}
 \frac{dU}{dt}  &=&
\frac{1}{8\pi^2}
 \Big\{6Uh^2_t + Dh^2_b - G_U m_{1/2}^2\Big\} \label{eq:28} \\
\frac{dD}{dt}  &=& \frac{1}{8\pi^2}
\Big\{Uh^2_t + 6Dh^2_b - G_D  m_{1/2}^2\Big\}  \label{eq:29}
\end{eqnarray}
where $G_U = G_Q + G_{H_2} + G_{U^c}$ and $G_D = G_Q + G_{H_1} + G_{B^c}$
To simplify the above coupled equations we make the following
transformations
\begin{eqnarray} \begin{tabular}{ccccc}
U  &=& $\tau $x  , $\tau $ &=& $\exp \Big\{-\frac{3}{4\pi^2}
 \int^t_{t_0}  h_t^2dt^\prime\Big\} $\label{eq:30} \\
D  &=& $\sigma $y  , $\sigma  $&=& $\exp \Big\{-\frac{3}{4\pi^2}
\int^t_{t_0}  h^2_b dt^\prime\Big\}$
\label{eq:30a}
\end{tabular}
\end{eqnarray}
Then Equs(\ref{eq:28}-\ref{eq:29}) can be written in the form
\begin{eqnarray}
\tau \frac{dx}{dt} &=& \frac{1} {6}  \frac{d\sigma}{dt} y -
\frac{G_U}{8\pi^2} m_{1/2}^2 \label{eq:31} \\
\sigma \frac{dy} {dt} &=& \frac{1} {6}  \frac{d\tau}{dt} x -
\frac{G_D}{8\pi^2} m_{1/2}^2
\label{eq:32}
\end{eqnarray}
It is trivial to check that D.E. (\ref{eq:31}) in the case $h_t \gg h_b$
can be solved independently giving the solution (\ref{eq:21}) for the
up-squarks and the higgs $H_2$.
Moreover, using the fact that $G_U\approx G_D \approx 2\, G_Q$,
one can easily obtain an exact solution of equs(\ref{eq:31}-\ref{eq:32}).
In the general case, (\ref{eq:31}-\ref{eq:32}) can combine to two simple,
second order differential equations  of the form
 \begin{eqnarray}
\frac{d^2}{dQ^2} y -  \frac {\alpha}{\beta} y =
 -\frac{d}{dQ}(\frac{g}{\beta}) - \frac{f}{\beta} \label{eq:33}\\
\frac{d^2}{dP^2} x -  \frac {\beta}{\alpha} x =
 -\frac{d}{dP}(\frac{f}{\alpha}) - \frac{g}{\alpha}
\label{eq:34}
\end{eqnarray}
with
\begin{eqnarray}
\alpha  &=& \frac{\sigma}{\tau} \frac{h^2_b}{8\pi^2} \equiv
\frac{1} {\tau}  \frac{d\sigma}{dt}   ,\qquad dP = \alpha dt
\label{eq:35}\\
\beta &=& \frac{\tau}{\sigma} \frac{h^2_t}{8\pi^2} = \frac{1}
{\sigma}  \frac {d\tau}{dt}  , \qquad dQ = \beta dt
\label{eq:36}
\end{eqnarray}
and
\begin{equation}
f(t)=\frac{G_U}{8\pi^2\tau}m_{1/2}^2 \, ,
\qquad  g(t)=\frac{G_D}{8\pi^2\sigma}m_{1/2}^2
\label{eq:37}
\end{equation}
These are both decoupled and have a standard second order form with a
non-homogeneous part on the R.H.S.
Since the coefficients are known functions of the analytic solutions
$h_b(t), h_t(t)$ they can be solved either numerically or in particular
cases by  standard methods yielding expressions for the sums of $m_{U_i},
m_{D_i}$ .  Then the expressions for the individual masses may be
obtained in the way described in ref\cite{leo}. Detailed results of the
above procedure will be presented elsewere.


{\bf Acknowledgements }
{\it We have been informed that
  C. Kounnas and F. Zwirner have obtained similar expressions
for $h_{t/b}$ couplings. We would like to thank them
 for discussions and communicating to us their results and,
  M. Carena and C. Wagner for
discussions on their work.  G.K.L. would like to acknowledge
a usefull discussion with I. Antoniadis.}

\end{document}